\documentclass[lettersize,journal]{IEEEtran}
\usepackage{amsmath,amsfonts,amssymb,amsthm}
\usepackage{xcolor}
\usepackage{algorithm}
\usepackage{algpseudocode}
\usepackage{array}
\usepackage{textcomp}
\usepackage{stfloats}
\usepackage{url}
\usepackage{verbatim}
\usepackage{graphicx}
\usepackage{cite}
\usepackage{multirow}
\usepackage{caption}
\usepackage{subcaption}
\usepackage{hhline}

\def\BibTeX{{\rm B\kern-.05em{\sc i\kern-.025em b}\kern-.08em
    T\kern-.1667em\lower.7ex\hbox{E}\kern-.125emX}}
\usepackage{balance}

\def\Re{\mathop{\mathrm{Re}}}

\def\b0{{\pmb{0}}} 

\begin{document}

\bstctlcite{IEEEexample:BSTcontrol}
	
\title{Efficient Solvers for Coupling-Aware Beamforming in Continuous Aperture Arrays}

\author{Geonhee Lee, Kwonyeol Park, Hyeongjun Park, Jinwoo An, and Junil Choi
\thanks{
        
        G. Lee, H. Park, J. An, and J. Choi are with the School of Electrical Engineering, Korea Advanced Institute of Science and Technology, Daejeon 34141, South Korea (e-mail: \{peppermint01, mika0303, freddy1, junil\}@kaist.ac.kr).

        K. Park is with the S.LSI Division, Samsung Electronics Company Ltd., Hwaseong-si, 18448, South Korea, and also with the School of Electrical Engineering, Korea Advanced Institute of Science and Technology, Daejeon 34141, South Korea (e-mail: kwon10.park@kaist.ac.kr).
}}
\maketitle

\begin{abstract}
In continuous aperture arrays (CAPAs), careful consideration of the underlying physics is essential, among which electromagnetic (EM) mutual coupling plays a critical role in beamforming performance. Building on a physically consistent mutual coupling model, the beamforming design is formulated as a functional optimization whose optimality condition leads to a Fredholm integral equation. The incorporation of the coupling model, however, substantially increases computational complexity, necessitating efficient and accurate integral equation solvers. In this letter, we propose two efficient solvers: 1) a coordinate-transformation-based kernel approximation that preserves the operator structure while alleviating discretization demands, and 2) a direct lower-upper (LU)-based solver that stably handles the Nyström-discretized system. Numerical results demonstrate improved accuracy and reduced computational overhead compared to conventional methods, with the LU-based solver emerging as an efficient and scalable solution for large-scale CAPA optimization via offline factorization.
\end{abstract}


\section{Introduction}
\label{Intro}
Continuous aperture arrays (CAPAs) are emerging as a promising electromagnetic (EM) architecture characterized by aperture-level continuity\cite{liu2025capa}. Such continuity relaxes the discrete spacing constraints of conventional arrays, allowing near-continuous manipulation of surface current distributions. Consequently, precise amplitude and phase control can be achieved across the aperture, enabling the array’s EM properties to be effectively exploited. Prior works have investigated various design strategies for CAPAs, including calculus-of-variations-based formulations to characterize energy-efficient current distributions \cite{CoV} and linear beamforming schemes to improve tractability in practical design \cite{linear_CAPA}. The weighted minimum mean-square error (WMMSE) framework enables sum-rate optimization in CAPA-enabled transceivers \cite{WMMSE_CAPA}. Together, these efforts highlight the need for careful modeling of EM properties in CAPA system design.

More recently, an EM coupling kernel for CAPAs and a corresponding numerical methodology have been proposed, laying the foundation for physically consistent CAPA beamforming \cite{mutual_CAPA}. Within this framework, two approaches have been proposed to solve the coupling-aware integral equation for optimal beamforming. Although both aim to solve the equation effectively, each exhibits notable limitations that affect stability or computational efficiency. Kernel approximation (KA) provides closed-form inverse operators that offer physical insight and fast computation, while explicit inversion can lead to numerical instability. Meanwhile, the conjugate gradient (CG) method avoids direct inversion through an iterative procedure. However, each iteration requires dense operators, and the iteration count grows with a problem size and an operating frequency, leading to high computational cost.

Incorporating mutual coupling into existing CAPA algorithms further intensifies these challenges. For example, WMMSE-based CAPA algorithms require multiple solves of the integral equation, making numerical stability and efficiency essential. Accordingly, high approximation accuracy is required even with coarse Gauss–Legendre (GL) sampling. At the same time, the need for repeated solutions of the integral equations necessitates low-complexity strategies capable of balancing accuracy, robustness, and efficiency.

In this letter, we develop two efficient computational strategies for solving the EM mutual coupling problem in CAPAs, with a focus on enabling fast and accurate approximation while supporting reliable numerical behavior. The main contributions of this letter are summarized as follows.
\begin{itemize}
\item We first develop a polar-trigonometric kernel approximation (PKA) that enables rapid, high-fidelity modeling while preserving analytical structure, supporting reliable coupling-aware continuous-aperture beamforming.

\item We also demonstrate that a direct lower-upper (LU)-based solver reliably resolves the Nyström-discretized system, achieving stable performance with reduced computational complexity.

\item Numerical results demonstrate that the proposed methods achieve superior accuracy and lower computational cost than conventional KA and CG methods.

\end{itemize}

Collectively, this work establishes practical and scalable techniques for coupling-aware beamforming, enabling numerically stable and low-latency optimization even for electrically large continuous apertures.

\section{System Model and Problem Formulation}
\label{sec2}

In this section, we establish the theoretical framework for the CAPA system. We begin by characterizing the EM radiation mechanism of a continuous aperture and the resulting mutual coupling effects. Based on these foundations, we formulate the array gain maximization problem and derive the structure of the optimal current distribution.

\subsection{CAPA and Radiated Field}
\label{sec2.A}
We consider a transmitter architecture based on CAPA, where the radiating interface is modeled as a continuous surface $\mathcal{S}$. 
The aperture is assumed to be a rectangular planar surface lying in the $xy$-plane, defined as \cite{mutual_CAPA, WMMSE_CAPA}
\begin{align}
    \mathcal{S} = \left\{ [s_x, s_y, 0]^\mathrm{T} \;\middle|\; |s_x| \le \frac{L_x}{2}, \; |s_y| \le \frac{L_y}{2} \right\},
\end{align}
where $s_x$ and $s_y$ denote the local spatial coordinates. $L_x$ and $L_y$ represent the physical width and height of the aperture surface $\mathcal{S}$, respectively.

According to \cite{degrees_of_freedom}, the radiated electric field $\mathbf{E}_{\mathrm{rad}}(\mathbf{r}) \in \mathbb{C}^{3}$  at an arbitrary observation point $\mathbf{r}$ is expressed as follows
\begin{align} \label{Green}
    \mathbf{E}_{\mathrm{rad}}(\mathbf{r})=\int_{\mathcal{S}} \mathbf{G}(\mathbf{r}-\mathbf{s})\mathbf{J}(\mathbf{s}) d\mathbf{s},
\end{align} 
where $\mathbf{G}(\mathbf{r}-\mathbf{s})$ denotes the free-space dyadic Green's function and $\mathbf{J}(\mathbf{s})$ represents the complex electric surface current density at location $\mathbf{s}$. As a linear operator, the dyadic Green's function maps the current source to the resulting radiated electric field and is defined as
\begin{align} \label{Green_definition}
\mathbf{G}(\mathbf{s}) = -j k_0 Z_0 
\left( \mathbf{I}_3 + \frac{1}{k_0^2} \nabla \nabla \right) 
\frac{e^{j k_0 \lVert \mathbf{s} \rVert_2}}{4\pi \lVert \mathbf{s} \rVert_2}.
\end{align}
Here, $\lambda$ is the wavelength, $Z_0 = 120\pi$ is the intrinsic impedance of free space, and $k_0 = 2\pi/\lambda$ is the wavenumber.

We consider a narrowband single-carrier system with a purely $y$-directed surface current over the transmit aperture, where the source current density is expressed as $\mathbf{J}(\mathbf{s}) = x(\mathbf{s})\mathbf{u}_y$ with $\mathbf{u}_y = [0,\,1,\,0]^{\mathrm T}$ denoting the unit vector along the $y$-axis and $x(\mathbf{s})$ representing the scalar current excitation at location $\mathbf{s}$. Let $y(\mathbf{r})$ denote the effective electric field at the receiver located at $\mathbf{r}$. Under line-of-sight (LoS) propagation and perfect polarization alignment between the transmitter and a single-antenna receiver, $y(\mathbf{r})$ is given by 
\begin{align}
    y(\mathbf{r}) &=\mathbf{u}_y^\mathrm{T} \mathbf{E}_{\mathrm{rad}}(\mathbf{r}) \nonumber\\
    &=\int_{\mathcal{S}}  \mathbf{u}_y^\mathrm{T} \mathbf{G}(\mathbf{r}-\mathbf{s}) \mathbf{J}(\mathbf{s}) \, d\mathbf{s} 
    = \int_{\mathcal{S}} h(\mathbf{r},\mathbf{s}) x(\mathbf{s}) \, d\mathbf{s},
 \end{align}
where $h(\mathbf{r},\mathbf{s}) = \mathbf{u}_y^\mathrm{T} \mathbf{G}(\mathbf{r}-\mathbf{s}) \mathbf{u}_y.$ In the far-field regime, with $R_0=\|\mathbf{r}\|_2$ greatly exceeds the aperture dimensions, the spherical wavefront can be approximated by a plane wave\cite{mutual_CAPA}. Under this approximation, the channel response depends only on $\mathbf{s}$ for a fixed receive location $\mathbf{r}$ and reduces to
\begin{align}
    h(\mathbf{s}) = \alpha(\boldsymbol{\kappa}_{\mathrm{r}})
    e^{-j \boldsymbol{\kappa}_{\mathrm{r}}^{\mathrm{T}} \mathbf{s}},
    \label{eq:plane_wave_channel}
\end{align}
where $\boldsymbol{\kappa}_{\mathrm{r}} = k_0[\sin\theta\cos\phi,\sin\theta\sin\phi]^\mathrm{T}$ with the elevation and azimuth angles of the propagation direction $\theta$ and $\phi$, respectively.
The complex gain $\alpha(\boldsymbol{\kappa}_r)$ is given by
\begin{equation}
    \alpha(\boldsymbol{\kappa}_r) = \frac{-j k_0 Z_0 e^{j k_0 R_0}}{4\pi R_0}
    \left( 1 - \sin^2 \theta \sin^2 \phi \right).
\end{equation}

\subsection{Mutual Coupling in CAPA Transmitters}
\label{sec2.B}
Mutual coupling in CAPAs refers to EM interactions whereby the power needed to support a current distribution is affected by both its self-generated fields and the fields from neighboring parts of the aperture. Based on the physical model established in \cite{mutual_CAPA}, the total electric field at the aperture surface $\mathbf{E}_{\mathrm{tot}}(\mathbf{s})$ is decomposed into dissipative and radiative components as
\begin{align}
    \mathbf{E}_{\mathrm{tot}}(\mathbf{s}) = \mathbf{E}_{\mathrm{diss}}(\mathbf{s}) + \mathbf{E}_{\mathrm{rad}}(\mathbf{s}).
\end{align}
The dissipated field, accounting for surface resistance $Z_s$, is modeled as $\mathbf{E}_{\mathrm{diss}}(\mathbf{s}) = Z_s \mathbf{J}(\mathbf{s})$. Consequently, the total field $\mathbf{E}_{\mathrm{tot}}(\mathbf{s})$ is given by \cite{mutual_CAPA}
\begin{align}
    \mathbf{E}_{\mathrm{tot}}(\mathbf{s}) = \int_{\mathcal{S}} \{Z_s \delta(\mathbf{s}-\mathbf{z}) \mathbf{I}_3 + \mathbf{G}(\mathbf{s} - \mathbf{z})\} \mathbf{J}(\mathbf{z}) d\mathbf{z}.
\end{align}
The average EM transmit power $P_{\mathrm{em}}$ is then evaluated by the following quadratic functional, expressed as \cite{pizzo_2025}
\begin{align}
\begin{split}
P_{\mathrm{em}} 
&= \frac{1}{2} \int_{\mathcal{S}} \int_{\mathcal{S}} 
\mathbf{J}^\mathrm{H}(\mathbf{s}) \\
&\quad \times \operatorname{Re}\{ 
Z_s \delta(\mathbf{s}-\mathbf{z}) \mathbf{I}_3 
+ \mathbf{G}(\mathbf{s}-\mathbf{z}) 
\}
\mathbf{J}(\mathbf{z}) \, d\mathbf{z} d\mathbf{s}.
\end{split}
\end{align}

For a purely $y$-directed surface current $\mathbf{J}(\mathbf{s})=x(\mathbf{s})\mathbf{u}_y$, the power functional reduces to the scalar form
\begin{align}
    P_{\mathrm{em}} = \frac{1}{2} \int_{\mathcal{S}} \int_{\mathcal{S}} x^*(\mathbf{s}) c(\mathbf{s}-\mathbf{z}) x(\mathbf{z}) d\mathbf{z} d\mathbf{s},
\end{align}
where the scalar coupling kernel $c(\mathbf{s})$ is given by
\begin{align}
    c(\mathbf{s}) &= Z_s \delta(\mathbf{s}) + \Re \left\{ \mathbf{u}_y^\mathrm{T} \mathbf{G}(\mathbf{s}) \mathbf{u}_y \right\} \nonumber\\
    &= Z_s \delta(\mathbf{s}) + k_0 Z_0 \left(1 + \frac{1}{k_0^2} \partial_y^2 \right) 
    \frac{\sin(k_0 \lVert \mathbf{s} \rVert_2)}{4\pi \lVert \mathbf{s} \rVert_2}.
    \label{coupling_kernel}
\end{align}
We define the second term as the radiation coupling component, such that 
$c(\mathbf{s}) = Z_s \delta(\mathbf{s}) + c_{\mathrm{rad}}(\mathbf{s})$.

\subsection{Problem Formulation and Optimal Beamforming Structure}
\label{sec2.C}
Following the framework established in \cite{mutual_CAPA}, we optimize the continuous current distribution $x(\mathbf{s})$ to maximize the array gain at the target receiver subject to a total transmit power constraint $P_t$. This leads to the following functional optimization problem:
\begin{align}
&\mathrm{(P1):}\,\, \underset{ x(\mathbf{s}) }{\max} \quad \left| \int_{\mathcal{S}} h(\mathbf{s}) x(\mathbf{s}) \, d\mathbf{s} \right|^2 \nonumber \\
&\quad \mathrm{s.t.}\quad \frac{1}{2} \int_{\mathcal{S}} \int_{\mathcal{S}} x^*(\mathbf{s}) c(\mathbf{s}-\mathbf{z}) x(\mathbf{z}) \, d\mathbf{z} d\mathbf{s} \le P_t. \tag{P1-a} 
\end{align}
By applying the calculus of variations to (P1), the optimal current distribution $x_{\mathrm{opt}}(\mathbf{s})$ is proportional to an auxiliary function $v(\mathbf{s})$, which satisfies the Fredholm integral equation:
\begin{align} \label{eq:v_solve}
    \int_{\mathcal{S}} c(\mathbf{s}-\mathbf{z}) v(\mathbf{z}) \, d\mathbf{z} = h^*(\mathbf{s}), 
    \quad \forall \mathbf{s} \in \mathcal{S}.
\end{align}
The optimal distribution is then given by
\begin{align}
    x_{\mathrm{opt}}(\mathbf{s})
    = \sqrt{ \frac{2P_t}{\int_{\mathcal{S}} h(\mathbf{z}) v(\mathbf{z}) \, d\mathbf{z}} } \, v(\mathbf{s}),
\end{align}
where scaling ensures the transmit power constraint.

Hence, the key computational task is to solve \eqref{eq:v_solve} to recover $v(\mathbf{s})$, 
as it directly determines the optimal current distribution $x_{\mathrm{opt}}(\mathbf{s})$. However, the continuous formulation renders the integral equation generally intractable, necessitating efficient solvers. 
The next section presents numerical strategies for this problem, focusing on accurate and computationally efficient solution methods.

\section{Coupling Integral Equation Solvers} \label{sec3}
This section briefly reviews conventional algorithms as a baseline and proposes two novel approaches: a polar–trigonometric quadrature for accurate KA and an LU decomposition method for the Nyström-discretized system.

\subsection{Review of the Conventional Kernel Approximation}

As described in \cite{mutual_CAPA}, the KA method aims to obtain a tractable representation of the coupling kernel $c(\mathbf{s})$ that enables a closed-form inverse. Specifically, the radiation mutual coupling kernel $c_{\mathrm{rad}}(\mathbf{s})$ is expressed in the wavenumber domain through the two-dimensional (2D) Fourier transform, yielding $C_{\mathrm{rad}}(\boldsymbol{\kappa})$, which is given by 
\begin{align}
    C_{\text{rad}}(\boldsymbol{\kappa}) = 
    \begin{cases} 
    \frac{Z_0(1 - \kappa_y^2/k_0^2)}{2\sqrt{1 - \|\boldsymbol{\kappa}\|_2^2/k_0^2}}, & \|\boldsymbol{\kappa}\|_2 \leq k_0, \\
    0, & \|\boldsymbol{\kappa}\|_2 > k_0,
    \end{cases}
    \label{eq:Forier_Transform}
\end{align}
where the wavenumber vector is defined as $\boldsymbol{\kappa} = [\kappa_x, \kappa_y, 0]^\mathrm{T}$.
The spatial kernel $c_{\text{rad}}(\mathbf{s})$ is then recovered via the inverse 2D Fourier transform as
\begin{align}
    c_{\text{rad}}(\mathbf{s}) = \frac{1}{(2\pi)^2} 
    \int_{-k_0}^{k_0} 
    \int_{-\sqrt{k_0^2 - \kappa_x^2}}^{+\sqrt{k_0^2 - \kappa_x^2}} 
    C_{\text{rad}}(\boldsymbol{\kappa}) e^{j \boldsymbol{\kappa}^\mathrm{T} \mathbf{s}} 
    d\boldsymbol{\kappa}. 
\end{align}

To discretize this integral, \cite{mutual_CAPA} employed Gauss--Legendre (GL) quadrature in Cartesian coordinates, where the sampling points are confined to the disk $\|\boldsymbol{\kappa}\|_2\le k_0$. Since the integration is two-dimensional, the approximation takes the form of a double summation:
\begin{align}
    c_{\text{rad}}(\mathbf{s}) \approx \sum_{n=1}^M \sum_{m=1}^M a_{nm} e^{j \boldsymbol{\kappa}_{nm}^\mathrm{T} \mathbf{s}}.
    \label{eq:GL_quadrature}
\end{align}
To derive the quadrature coefficient $a_{nm}$, first let $(\theta_n,\omega_n)$ represent the Gauss--Legendre node--weight pairs on $[-1,1]$. To construct wavenumber samples $(\kappa_{x,n}, \kappa_{y,nm})$ within the disk $\|\boldsymbol{\kappa}\|_2\le k_0$, we discretize the $x$-component as
\begin{align}
    \kappa_{x,n} = k_0 \theta_n, 
    \qquad
    w_{x,n} = k_0 \omega_n,
    \label{eq:kappaW_x_GL}
\end{align}
where $\kappa_{x,n}\in[-k_0,k_0]$ is the wavenumber sample with quadrature weight $w_{x,n}$. For each $\kappa_{x,n}$, the range of $\kappa_{y,nm}$ is given by $\left[-\sqrt{k_0^2-(\kappa_{x,n})^2},\,\sqrt{k_0^2-(\kappa_{x,n})^2}\right].$ With a second GL quadrature, the y-component is sampled as $
    \kappa_{y,nm}
    = \sqrt{k_0^2-(\kappa_{x,n})^2}\,\theta_m,
    \,
    w_{y,nm} 
    = \sqrt{k_0^2-(\kappa_{x,n})^2}\,\omega_m.$ The associated quadrature coefficient is then given by
\begin{align}
    a_{nm}
    = \frac{w_{x,n} w_{y,nm}}{(2\pi)^2}
      C_{\mathrm{rad}}\!\left(\boldsymbol{\kappa}_{nm}\right),
    \label{eq:rho_GL}
\end{align}
where $\boldsymbol{\kappa}_{nm}=[\kappa_{x,n}\;\kappa_{y,nm}\;0]^\mathrm{T}$.

By utilizing \eqref{coupling_kernel} and \eqref{eq:GL_quadrature}, $c(\mathbf{s})$ can be approximated as
\begin{align}
    c(\mathbf{s}) \approx Z_s \delta(\mathbf{s}) + \sum_{n=1}^M \sum_{m=1}^M a_{nm} e^{j \boldsymbol{\kappa}_{nm}^\mathrm{T} \mathbf{s}}.
    \label{eq:kernel_expansion}
\end{align}
In principle, \eqref{eq:v_solve} can be solved via the inverse kernel $c^{-1}(\mathbf{s})$, defined by the orthogonality condition $\int_{\mathcal{S}} c^{-1}(\mathbf{z}'-\mathbf{s})\,c(\mathbf{s}-\mathbf{z})\,d\mathbf{s} = \delta(\mathbf{z}'-\mathbf{z})$ \cite{mutual_CAPA}. This yields the formal solution $v(\mathbf{z}) = \int_{\mathcal{S}} c^{-1}(\mathbf{z}-\mathbf{s})\,h^*(\mathbf{s})\,d\mathbf{s}$. When applied to the kernel $c(\mathbf{s})$, the inverse kernel admits the representation
\begin{align}
    c^{-1}(\mathbf{z}-\mathbf{s}) = \frac{1}{Z_s} \delta(\mathbf{z}-\mathbf{s}) + \sum_{n=1}^M \sum_{m=1}^M b_{nm} e^{j \boldsymbol{\kappa}_{n}^\mathrm{T} \mathbf{z}-j\boldsymbol{\kappa}_{m}^\mathrm{T} \mathbf{s}},
    \label{eq:inverse_kernel}
\end{align}
where $b_{nm}$ are chosen to satisfy the orthogonality condition. Therefore $v(\mathbf{s})$ is given by
\begin{align}
    v(\mathbf{z}) = \frac{1}{Z_s} h^*(\mathbf{z}) + \sum_{n=1}^M \sum_{m=1}^M b_{nm} e^{j\boldsymbol{\kappa}_{n}^\mathrm{T} \mathbf{z}} 
    \left( \int_{\mathcal{S}} e^{-j \boldsymbol{\kappa}_{m}^\mathrm{T} \mathbf{s}} h^*(\mathbf{s}) \, d\mathbf{s} \right).
    \label{eq:v_structured}
\end{align}
This closed-form approximation provides a tractable inverse operator for scalable beamforming design in CAPA systems.

\subsection{Proposed Method I: Polar-Trigonometric Kernel Approximation}
While the Cartesian GL quadrature in \eqref{eq:rho_GL} is straightforward, it becomes numerically unstable near the boundary of the propagation region.
As $\|\boldsymbol{\kappa}\|_2 \to k_0$, the denominator in (\ref{eq:Forier_Transform}) approaches zero, causing $C_{\mathrm{rad}}(\boldsymbol{\kappa})$ to diverge. Although the case $\kappa_y = k_0$ remains finite due to the simultaneous vanishing of numerator, the integrand is otherwise unbounded near the boundary, leading to large integration errors and requiring prohibitively large $M$ for accurate evaluation.

To mitigate this numerical instability, we adopt a polar coordinate transformation combined with a trigonometric substitution, expressing the wavenumber components as
\begin{align}
    \kappa_x = k_0 \sin \theta \cos \phi, \quad
    \kappa_y = k_0 \sin \theta \sin \phi,
    \label{eq:coordinate}
\end{align}
where $\theta \in [0, \pi/2]$ and $\phi \in [0, 2\pi]$. Substituting these expressions into the denominator yields
\begin{align}
    \sqrt{1 - \|\boldsymbol{\kappa}\|_2^2/k_0^2}
    = \sqrt{1 - \sin^2\theta}
    = \cos\theta.
\end{align}
Since the Jacobian of this transformation is $k_0^2 \sin\theta \cos\theta$, the $\cos\theta$ term in the denominator is exactly canceled, thereby removing the singular behavior when $\|\boldsymbol{\kappa}\|_2 \to k_0$. Under the polar--trigonometric transformation, the coefficients $a_{nm}$ are modified as $\tilde{a}_{nm}$, given by
\begin{align}
    \tilde{a}_{nm}
    = \frac{Z_0 k_0^2}{2(2\pi)^2}
    \left(1 - \sin^2\theta_n \sin^2\phi_m\right)
    \sin\theta_n \, w_{\theta,n} w_{\phi,m}.
\end{align}

Let $\{(\xi_n,\omega_n)\}_{n=1}^{M}$ denote the GL nodes and weights on the reference interval $[-1,1]$.
For the elevation angle, the nodes $\theta_n$ and weights $w_{\theta,n}$ are obtained by linearly mapping this interval onto $[0,\pi/2]$:
\begin{align}
    \theta_n = \frac{\pi}{4}(\xi_n + 1), \quad
    w_{\theta,n} = \frac{\pi}{4}\omega_n, \quad n=1,\dots,M.
    \label{eq:elevation}
\end{align}
Likewise, the azimuth nodes $\phi_m$ and weights $w_{\phi,m}$ are mapped onto $[0,2\pi]$ as
\begin{align}
    \phi_m = \pi(\xi_m + 1), \quad
    w_{\phi,m} = \pi\omega_m, \quad m=1,\dots,M.
    \label{eq:azimuth}
\end{align}

By combining (\ref{eq:coordinate}), (\ref{eq:elevation}), and (\ref{eq:azimuth}), the sampled wavenumber vectors ${\boldsymbol{\kappa}}_{nm}$ for the proposed method are obtained as
\begin{align}
    \boldsymbol{\kappa}_{nm} =
    \begin{bmatrix}
    k_0 \sin \theta_n \cos \phi_m &
    k_0 \sin \theta_n \sin \phi_m &
    0
    \end{bmatrix}^{\mathrm{T}}.
\end{align}
This sampling ensures that the $M^2$ points follow the Gauss--Legendre rule over the disk, while the trigonometric substitution handles the kernel singularity at $\theta_n=\pi/2$.

\subsection{Proposed Method II: Direct Numerical solve via LU Decomposition}

According to (\ref{coupling_kernel}) and (\ref{eq:v_solve}), the optimal auxiliary function $v(\mathbf{s})$ is obtained by solving the integral equation:
\begin{align} \label{eq:Fredholm}
    \int_{\mathcal{S}} c_{\mathrm{rad}}(\mathbf{s}-\mathbf{z}) v(\mathbf{z}) d\mathbf{z}
    + Z_s v(\mathbf{s})
    = h^*(\mathbf{s}), 
    \quad \forall \mathbf{s} \in \mathcal{S}.
\end{align}
The Nystr\"om method, a well-established technique for solving integral equations\cite{integral_equation}, is employed to enable numerical computation. Specifically, we discretize the spatial domain $\mathcal{S}$ using GL quadrature. Let $\{\mathbf{s}_n, w_n\}_{n=1}^{N}$ denote the quadrature nodes and weights, where $N=M^2$ is the total number of samples obtained from an $M \times M$ grid. Applying this discretization converts \eqref{eq:Fredholm} into a finite-dimensional linear system:
\begin{align}
    \sum_{m=1}^{N} 
    c_{\mathrm{rad}}(\mathbf{s}_n-\mathbf{s}_m)\, v(\mathbf{s}_m) w_m
    + Z_s v(\mathbf{s}_n) = h^*(\mathbf{s}_n), 
\end{align}
where $n=1,\dots,N.$ This can be expressed in matrix-vector form as
$\mathbf{C}\mathbf{v} = \mathbf{b}$, where $\mathbf{v} = [v(\mathbf{s}_1), \dots, v(\mathbf{s}_N)]^\mathrm{T}$ and $\mathbf{b}=[h^*(\mathbf{s}_1), \dots, h^*(\mathbf{s}_N)]^\mathrm{T}$. The matrix $\mathbf{C} \in \mathbb{C}^{N \times N}$ is defined by
\begin{align}
    [\mathbf{C}]_{nm} = c_\mathrm{rad}(\mathbf{s}_n - \mathbf{s}_m)w_m + Z_s \delta_{nm},
\end{align}
where $\delta_{nm}$ denotes the Kronecker delta.
In practice, the coupling matrix $\mathbf{C}$ tends to become ill-conditioned, particularly at high sampling densities and for small surface impedance values $Z_s$. As a result, directly computing the inverse of $\mathbf{C}$ may significantly magnify perturbations in the data, leading to severe loss of numerical accuracy and amplified  round-off errors.

To enhance numerical robustness, we instead solve the linear system $\mathbf{C}\mathbf{v}=\mathbf{b}$ using LU decomposition with partial pivoting. Specifically, the matrix is factorized as $\mathbf{P}\mathbf{C}=\mathbf{L}\mathbf{U}$, where $\mathbf{P}$ is a permutation matrix, $\mathbf{L}$ is a unit lower triangular matrix, and $\mathbf{U}$ is an upper triangular matrix. The overall procedure is summarized as follows\cite{matrix_computations}:
\begin{enumerate}
    \item \textbf{LU factorization:}\\
    Compute the factorization $\mathbf{P}\mathbf{C}=\mathbf{L}\mathbf{U}$ using partial pivoting to improve numerical stability.
    
    \item \textbf{Forward substitution:}\\
    Solve $\mathbf{L}\mathbf{y}=\mathbf{P}\mathbf{b}$ by evaluating the entries of $\mathbf{y}$ from top to bottom.
    
    \item \textbf{Backward substitution:}\\
    Solve $\mathbf{U}\mathbf{v}=\mathbf{y}$ by evaluating the entries of $\mathbf{v}$ from bottom to top.
\end{enumerate}

This approach avoids explicit matrix inversion by solving the system through stable triangular operations, thereby improving numerical stability and accuracy. 
To further enhance robustness, partial pivoting is incorporated during the factorization to control element growth and reduce sensitivity to round-off errors. 
Once the LU factors are computed, each solve requires only forward and backward substitutions, resulting in efficient computation with predictable complexity \cite{matrix_computations}. 

Meanwhile, \cite{mutual_CAPA} also proposed a CG method to solve (\ref{eq:v_solve}) without explicitly forming the inverse kernel. By reformulating the problem as a quadratic functional minimization, the CG method achieves stable convergence via the iterative updates as $
\mathbf{v}_{(n+1)} = \mathbf{v}_{(n)} + \alpha_{n}\mathbf{p}_{(n)}$ and $ 
\mathbf{p}_{(n+1)} = \mathbf{r}_{(n+1)} + \xi_{n}\mathbf{p}_{(n)},$ where $\mathbf{v}_{(n)}$, $\mathbf{r}_{(n)}$, and $\mathbf{p}_{(n)}$ denote the current iterate, residual, and conjugate search direction, respectively. The parameters $\alpha_n$ and $\xi_n$ denote the step size and update coefficient.

As shown in Fig.~5 of \cite{mutual_CAPA}, the required iteration count for the CG method grows with the operating frequency, increasing the computational burden. Since each iteration entails a dense matrix--vector multiplication with $\mathcal{O}(N^2)$ complexity, the overall computational cost becomes substantial.

In contrast, the LU-based solver shifts most of the computational effort to a one-time factorization of the coupling matrix. After this preprocessing step, subsequent solutions can be obtained with low overhead, making the approach well suited for real-time beamforming under fixed hardware configurations while maintaining strong numerical reliability.

\section{Simulation Results}\label{sec4}
In this section, we present the numerical results to evaluate the performance of proposed solvers for CAPA beamforming. Following \cite{mutual_CAPA}, we consider a rectangular planar aperture $\mathcal{S}$ with side lengths $L_x = L_y = 0.5$~m and set the surface resistance to $Z_s = 0.0128~\Omega$. The channel is modeled under a far-field LoS assumption, where the receiver is positioned at a distance $R_0 = 50$~m with elevation $\theta = 0$ and azimuth $\phi = 0$.

To assess the effectiveness of the proposed approaches, we compare the following four solvers:
\begin{enumerate}
    \item \textbf{KA}: The conventional kernel approximation method based on GL quadrature in Cartesian coordinates\cite{mutual_CAPA}.
    
    \item \textbf{CG}: The conjugate gradient method presented in \cite{mutual_CAPA}, serving as an iterative baseline for solving the Fredholm equation.
    
    \item \textbf{PKA (Polar-Trigonometric KA) [Proposed]}: The proposed solver employing a polar coordinate transformation with trigonometric substitution to KA method.
    
    \item \textbf{LU [Proposed]}: The proposed numerical solver of the Nyström-discretized system using LU decomposition.
\end{enumerate}

The primary performance metric is the normalized array gain, defined as \cite{mutual_CAPA}
\begin{align}
    G_{\mathrm{norm}} 
    = \frac{1}{P_\mathrm{t}}\left|\int_{\mathcal{S}} h(\mathbf{s})\, x(\mathbf{s}) \, d\mathbf{s}\right|^2,
\end{align}
which represents the ratio between the received signal power and the total transmit power $P_\mathrm{t}$.

\begin{figure}[t]
	\centering
	\includegraphics[width=1\columnwidth]{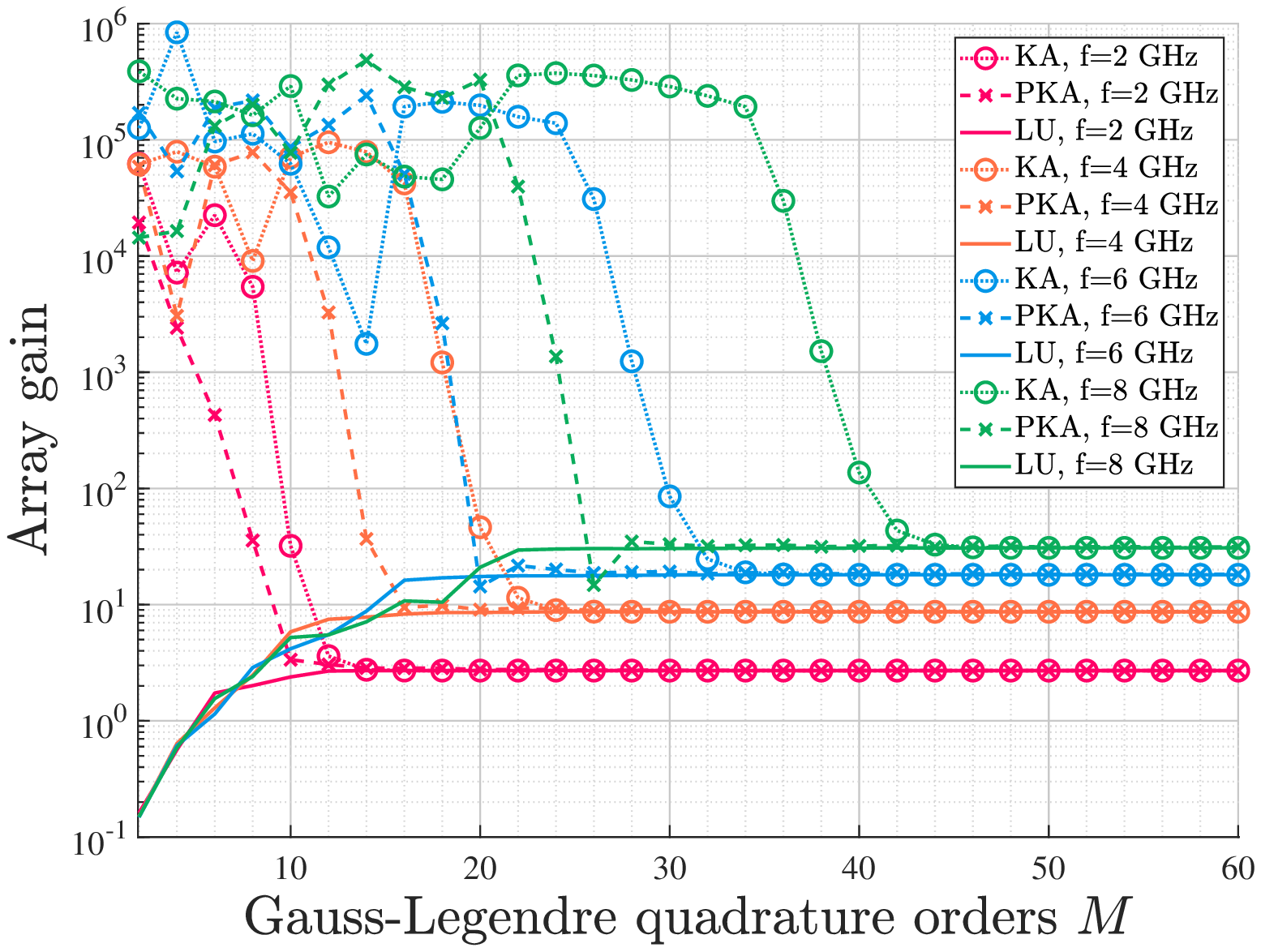}
	\caption{Normalized array gain versus Gauss--Legendre quadrature order $M$ for $L_x = L_y = 0.5~\mathrm{m}$.}
    \label{fig:convergence}
\end{figure}

\subsection{Approximation Performance and Numerical Stability}
\label{sec4.A}
In this subsection, we assess how accurately the proposed solvers approximate the true solution and examine their numerical stability in comparison with conventional benchmarks. Fig.~\ref{fig:convergence} shows the normalized array gain as a function of the number of GL quadrature points, denoted by $M$. Because the KA and PKA algorithms exhibit oscillatory behavior between even and odd orders, the results are sampled at intervals of two to improve visual clarity. As the CG and LU curves are nearly indistinguishable over the entire range of $M$, the CG curve is omitted.

The conventional KA method approaches the true value but exhibits irregular and oscillatory convergence. This behavior stems from the singularity of the coupling kernel near the propagation boundary, which amplifies integration errors under Cartesian coordinates. As a result, dense sampling is required to obtain a stable solution, leading to increased computational burden at higher operating frequencies.

In contrast, the proposed PKA method rapidly approximates the true value, attaining the optimal array gain with substantially fewer quadrature points. For example, at $f=8$~GHz, the required order for KA is 44 while that for PKA is 28, effectively shrinking the problem size from roughly $2000\times2000$ to $700\times700$. This improvement validates the effectiveness of the polar--trigonometric transformation in removing the kernel singularity. 

The LU solver exhibits highly stable convergence to the true array gain once sufficient spatial discretization is applied. This behavior is in sharp contrast with the KA-based methods, which exhibit less regular and oscillatory convergence, thereby highlighting the substantial advantage of the direct solver in numerical stability. Since the LU factorization can be performed offline and reused for different channel realizations, the direct solver is particularly attractive for practical systems requiring both accuracy and real-time operation.

Overall, the results confirm that the proposed methods substantially improve numerical stability while reducing computational effort compared to the conventional benchmark.

\subsection{Computational Complexity and Runtime}
\label{sec4.B}
Table~I compares the computational performance of the LU solver and the iterative CG method across different frequencies. All results are obtained with $M=25$. The LU decomposition is performed once and reused for $625$ solves with the LU solver (via forward and backward substitution), while the CG solver independently performs $625$ linear solves.

A key observation is the consistent runtime of the LU-based solve, with the total execution time remaining nearly constant at approximately $1.9$ sec despite increasing frequency. This stability stems from the fact that the problem matrix is determined solely by the aperture geometry and operating frequency, enabling the LU factors to be precomputed and reused. As a result, subsequent solves require only forward and backward substitutions, leading to predictable and low latency.

In contrast, the CG method demonstrates strong sensitivity to frequency. Its total runtime increases from $18.5$ sec at $2$ GHz to over $58.4$ sec at $6$ GHz, primarily driven by the growth in the average iteration count from $265.9$ to $877$. Since each CG iteration involves a large-scale dense matrix--vector multiplication, higher frequencies worsen the numerical conditioning and thereby increase the computational cost.

In conclusion, the direct LU solver delivers superior speed and scalability, making it particularly well suited for high-frequency CAPA beamforming scenarios.

\begin{table}[!t]
\centering
\caption{Computational complexity comparison between LU-based and CG-based solvers.}
\label{tab:LU_vs_CG}
\renewcommand{\arraystretch}{1.1}
\setlength{\tabcolsep}{4.5pt}
\begin{tabular}{|c|c|c|c|c|}
\hline
\textbf{Frequency} & \textbf{2~$\mathrm{GHz}$} & \textbf{4~$\mathrm{GHz}$} & \textbf{6~$\mathrm{GHz}$} & \textbf{8~$\mathrm{GHz}$} \\
\hline 
\textbf{LU decomp (1x)~[$\mathrm{ms}$]} & 652.584 & 625.216 & 703.268 & 669.988 \\
\hline
\textbf{LU solve (total)~[$\mathrm{ms}$]} & 1909.334 & 1904.806 & 1901.497 & 1880.552 \\
\hline
\textbf{CG solve (total)~[$\mathrm{ms}$]} & 18525.238 & 40852.152 & 58448.329 & 56571.415 \\ \hline
\textbf{CG iteration (avg.)} & 265.9 & 590.0 & 877.0 & 880.0 \\ \hline
\end{tabular}
\end{table}

\section{Conclusion}
In this letter, we addressed numerical challenges in continuous-aperture beamforming arising from the explicit modeling of EM mutual coupling. We proposed a coordinate-transformed kernel approximation that preserves the analytical structure while alleviating the need for excessively fine discretization. In addition, we showed that direct LU solver provides a reliable and numerically stable solution to the Nystr\"om-discretized system under Gauss--Legendre quadrature. Simulation results showed that the polar--trigonometric kernel approximation achieves faster and more consistent convergence to the optimal array gain with significantly fewer quadrature points than a benchmark, whereas the direct LU solver provides highly robust numerical stability along with predictable runtime. Moreover, offline factorization supports low-latency updates, making the proposed LU solver an efficient and scalable solution for large-scale CAPA optimization.

% \begin{figure}[!t]
%     \centering
%     \includegraphics[width=.65\columnwidth]{figures/NMSE_INT_Pwr_log_scale.eps}
%     \caption{NMSE comparison versus impulsive interference power~ratio.}
%     \label{fig:simulResult_intPwr}
% \end{figure}

%\input{Appendix}

\bibliographystyle{IEEEtran}
\bibliography{reference}

\vfill
	
\end{document}